\documentclass[10pt]{article}
\usepackage{epsf}
\usepackage{latexsym}
\title{\large \bf Spin-Galvanic Effect in Quantum Wells}
\author{ S.~D. Ganichev\,$^{1}$, E.~L.~Ivchenko\,$^2$, V.~V.~Bel'kov\,$^2$, S.A.~Tarasenko\,$^2$,
M.~Sollinger\,$^1$, \\ D.~Schowalter\,$^1$,
D.~Weiss\,$^1$,
W.~Wegscheider\,$^1$,
W.~Prettl\,$^1$\\
}

\date{\normalsize
$^1$ Fakult\"{a}t f\"{u}r Physik, Universit\"{a}t Regensburg, D-93040 Regensburg,
Germany
\\$^2$ A.F.~Ioffe Physico-Technical Institute, 194021 St. Petersburg,
Russia
 }

\begin{document}
\maketitle
\thispagestyle{empty}
{\bf Abstract:}

It is shown that a homogeneous non-equlibrium spin-polarization
in semiconductor
heterostructures results in  an electric current.
The microscopic origin of the effect is an inherent asymmetry of spin-flip
scattering in systems with lifted spin degeneracy caused by
{\boldmath$k$}-linear terms in the Hamiltonian.
%
%
\section{{Introduction}}
Much current interest  of condensed matter
physics is directed towards the understanding of various
manifestations of spin dependent phenomena. In particular, the
spin of electrons and holes in solid state systems is the decisive
ingredient for active spintronic devices~\cite{Wolf}. Here we
report on a new property of the spin-polarized
electron gas: its ability to drive an electric
current~\cite{Nature02}. While electrical currents are usually
generated by gradients of the potential, the carrier concentration
or the temperature, it is shown that a uniform non-equilibrium
spin orientation gives rise to an electric current.
This new spin-related phenomenon named spin-galvanic effect
has been observed recently in zinc-blende GaAs quantum well structures~\cite{Nature02}.
The microscopic origin of the effect observed in
low-dimensional electron systems
is an inherent asymmetry of the spin-flip scattering of electrons
in systems
with removed spin degeneracy of the band structure due to
{\boldmath$k$}-linear terms in the Hamiltonian.
Here we report on the investigation of spin-galvanic effect in InAs and GaAs QWs.
%
\begin{figure}[h]
\begin{center}
\mbox{\epsfxsize = 4cm \epsfbox{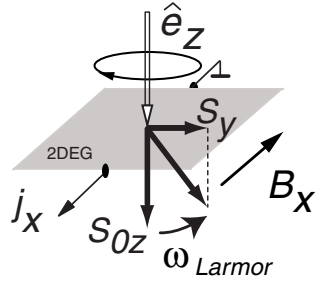}} \\
\end{center}
\caption{Experimental procedure to obtain a uniform spin polarization
in the plane of a QW. Electron spins are oriented normal to the plane
by circularly polarized radiation and rotated
into the plane by Larmor precession in a magnetic field.}
\label{F1}
\end{figure}
\section{{Overview}}
A basic symmetry property of  low dimensional zinc-blende based
structures is that they belong
to gyrotropic crystal classes. This means that
an axial vector of an average  spin polarization {\boldmath$S$}
and   a polar vector of  an electric current {\boldmath$j$}
may be linked by
a second rank pseudotensor {\boldmath$Q$}:
\begin{equation} j_{\alpha} =
\sum_{\beta}Q_{\alpha \beta} S_{\beta}.
\end{equation}
Non-zero
components of $Q_{\alpha\beta}$ can exist in QWs in contrast to the
corresponding bulk crystals.  In  (001)-grown QWs of
$C_{2v}$ symmetry only two linearly independent components,
$Q_{xy}$ and $Q_{yx}$, are different from zero ($x \parallel
[1\bar{1}0]$ and $y
\parallel [110]$).
Hence, to observe a spin polarization driven current a spin
component lying in the plane of the QW is required
(e.g. $S_y$ in Fig.\,\ref{F1}).
%
\begin{figure}[!h]
\begin{center}
\mbox{\epsfxsize = 6cm \epsfbox{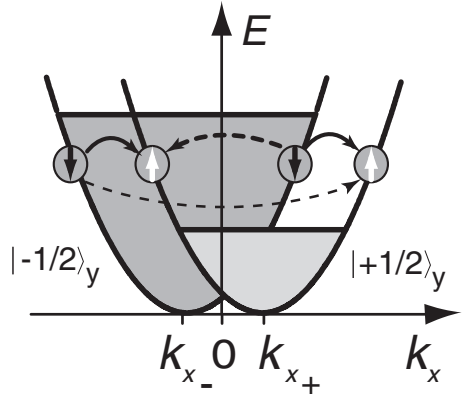}} \\
\end{center}
\caption{Sketch of spin orientation (shading)
at spin split subbands and current due to wavevector
dependent asymmetric spin-flip scattering.}
\label{F2}
\end{figure}
%
Microscopically, the spin-galvanic effect is caused by the asymmetric
spin-flip relaxation of spin polarized electrons in systems with
{\boldmath$k$}-linear contributions to the effective Hamiltonian.
The
lifting of spin degeneracy of 2DEG depicted in Fig.~\ref{F2} is a
consequence of a contribution to the Hamiltonian of the form $\hat
{H}_{{\bf k}} = \sum_{\gamma\alpha}\beta_{\alpha\gamma} \sigma_\alpha
k_\gamma$ where $\sigma_\alpha$ are the Pauli spin matrices
and {\boldmath$\beta$}
is a pseudotensor  subjected to the same symmetry restriction as {\boldmath$Q$} used
in Eq.~(1).
Fig.~\ref{F2} sketches the electron energy spectrum with the
$\beta_{yx}\sigma_y k_x$ term included.
This term leads to the splitting of the band into two branches with the
spin states $|\pm1/2\rangle_y$ relatively shifted along the
$k_x$-direction.
Spin orientation in $y$-direction yields an
unbalanced population in the spin-down and spin-up subbands. The current
flow is caused by the {\boldmath$k$}-dependent spin-flip relaxation
processes. Spins oriented in $y$-direction are scattered along $k_x$ from
the higher filled, e.g., spin-down subband, $|-1/2 \rangle_y$, to the less
filled spin-up subband, $|+1/2 \rangle_y$. Four quantitatively different
spin flip scattering events exist and are sketched in Fig.~\ref{F2} by
bent arrows. The spin-flip scattering rate depends on the values of the
wavevectors of the initial $k_{xi}$ and the final $ k_{xf}$ states,
respectively~\cite{Golub2}. Therefore spin-flip transitions, shown by solid
arrows in Fig.~\ref{F2}, have the same rates. They preserve the symmetric
distribution of carriers in the subbands and, thus, do not yield a current.
However, the two scattering processes shown by broken arrows are
inequivalent and generate an asymmetric carrier distribution around the
subband minima in both subbands. This asymmetric population results in a
current flow along the $x$-direction. Within our model of elastic
scattering the current is not spin polarized since the same number of
spin-up and spin-down electrons move in the same direction with the same
velocity.
The uniformity of the spin polarization
in space is preserved during the scattering processes.
\section{{Experimental results and discussion}}
To achieve a  homogeneous  non-equlibrium  spin polarization in experiment
we used optical spin
orientation.
Fig.\,\ref{F1} shows the geometry of the experiment. At normal incidence
of circularly polarized radiation the optical excitation yields a steady-state spin
orientation $S_{0z}$ in the $z$-direction.
To
obtain an in-plane component of the spins, necessary for the novel
effect described here, a
magnetic field, $B$,  was applied
(Fig.\,\ref{F1}). The field perpendicular to both the light
propagation direction $z$ and the optically oriented
spins, rotates the spins into the plane of the 2DEG due to Larmor
precession. With the magnetic field oriented along the $x$-axis we
obtain a non-equilibrium spin polarization $S_y$ which is
\begin{equation}
S_y = -\frac{\omega_L\tau_{s \perp}}{1 + (\omega_L
\tau_{s})^2}\:S_{0z}
\end{equation}
where $\tau_s = \sqrt{\tau_{s\parallel} \tau_{s\perp} }$ and
$\tau_{s \parallel}, \tau_{s \perp}$ are the longitudinal and
transverse electron spin relax\-ation times, $\omega_L$
is the Larmor frequency. Utilizing
the Larmor precession
we prepared the situation sketched in Fig.\,\ref{F1} where the spin polarization
$S_y$ lies in the plane. The denominator in Eq.\,2 yielding the
decay of $S_y$ for $\omega_L$ exceeding the inverse spin
relaxation
time
is well known from the Hanle effect~\cite{Meier}.

The experiments were carried out at various temperatures from room
temperature down to 4.2\,K
on $n$-GaAs single QWs of 7\,nm and 15\,nm width,
on $n$-GaAs single heterojunctions and on a single $n$-InAs QW of 15\,nm
or  7.6\,nm width.
These (001)-oriented samples
grown by molecular-beam-epitaxy  contain 2DEG systems with
electron  densities $n_s \simeq 2 \cdot 10^{11}$\,cm$^{-2}$ and
mobilities $\mu$ above $10^6$\,cm$^2$/\,Vs at T=4.2\,K.
Two pairs of point contacts were centered
on opposite sample edges along the direction $x \parallel
[1\bar{1}0]$ and $y \parallel
[110]$.
Two additional pairs of ohmic
contacts have been formed in the corners of the sample
corresponding to the $\langle 100 \rangle$ crystallographic
directions.
The radiation  of a cw Ti:Sapphire laser
at the wavelength 0.77\,$\mu$m was  applied for interband excitation.
As radiation source for intraband excitation a pulsed TEA-CO$_2$ laser
and a TEA-CO$_2$ laser pumped molecular far-infrared (FIR) laser were used.
Depending on the photon energy and quantum well band structure
the infrared  and FIR radiation induce direct optical transitions between
size quantized subbands
or, at longer wavelength, indirect optical transitions in the lowest subband.
%
\begin{figure}[h]
\begin{center}
\mbox{\epsfxsize = 7cm \epsfbox{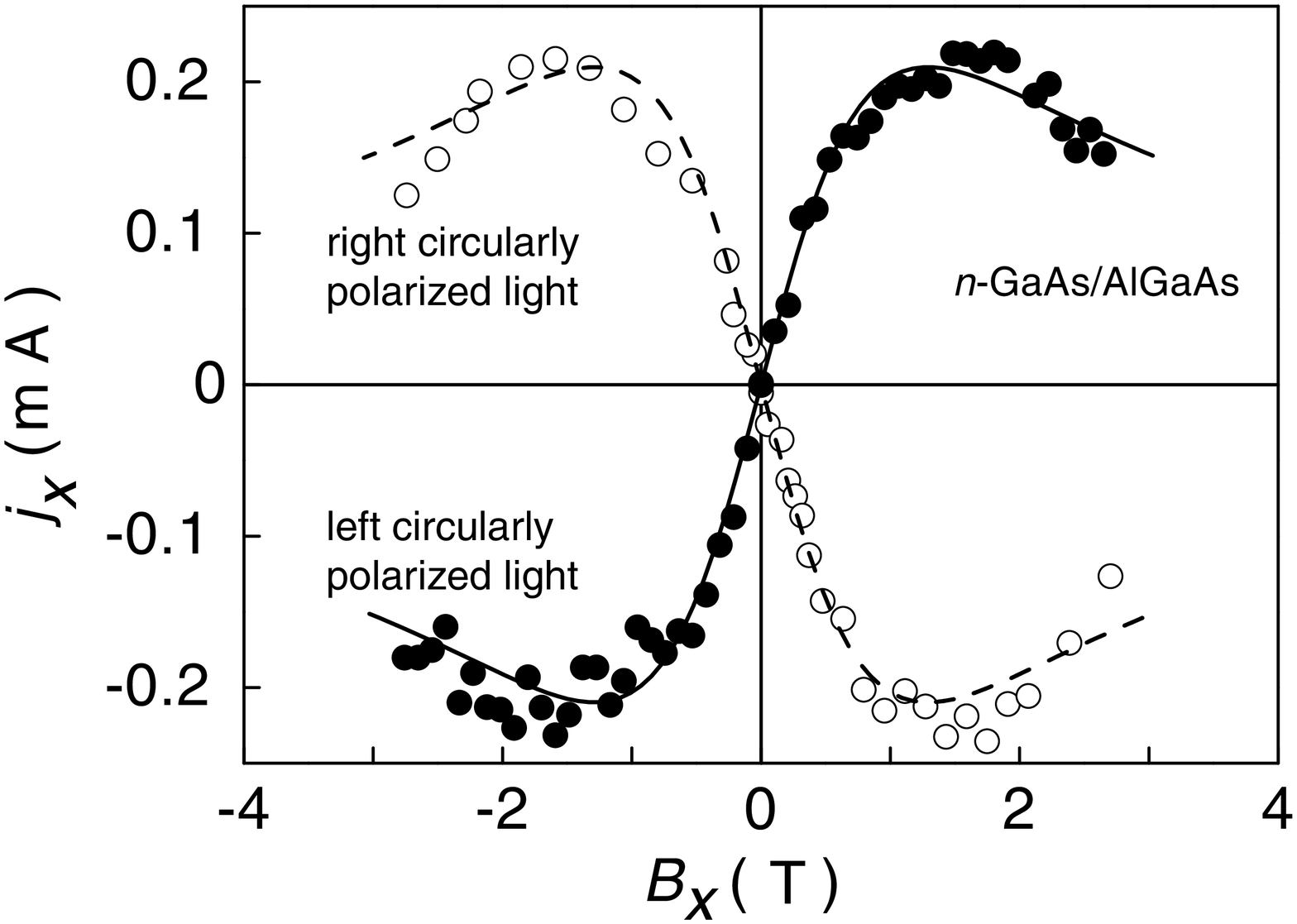}} \\
\end{center}
\caption{Current versus magnetic field obtained for an
$n$-GaAs/AlGaAs single heterojunction 
at $T=4.2$\,K and $\lambda = 148\,\mu$m. Curves are after Eq.\,(2).
}
\label{F3}
\end{figure}
Both for visible and infrared radiation a current has been observed for all
(001)-oriented $n$-type GaAs and InAs samples after applying an in-plane
magnetic field in the whole temperature range.
In Fig.~\ref{F3} the observed current as a function of the magnetic field
is shown for right and left handed circular polarization of $\lambda =
148\,\mu$m radiation. The polarity of the current depends on the
initial orientation of the excited spins and on the direction of the applied
magnetic field.

The current is parallel
(anti-parallel) to the magnetic field vector and follows the field as it is
rotated around the growth axis which has been checked by using of
different pairs of contacts.
For higher magnetic fields the current assumes a maximum and decreases upon
further increase of $B$.
This is ascribed to the Hanle effect, Eq.\,(2).
The  observation of the Hanle effect demonstrates that free carrier intra-subband
transitions can polarize
the spins of electron
systems.
In a very direct way the measurements allow to
obtain the spin relaxation time $\tau_s$ from the
peak position of the photocurrent where $\omega_L\tau_s = 1$.

The experiments demonstrate that in zinc-blende QWs
a spin polarization uniform in space
results in an electric current.
Therefore the spin-galvanic effect differs  from
other experiments where the spin current is caused by
gradients of potentials, concentrations etc. like
the spin-voltaic
effect~\cite{Silsbee,Sarma},
which, as the photo-voltaic effect, occurs in
inhomogeneous samples,
as well as from surface currents  induced by inhomogeneous spin
orientation~\cite{Dyakonov}.
\section{Spin-galvanic versus photogalvanic effect}
In this section we would like to point out the difference between the
spin-galvanic effect and another spin-related effect occuring in
zinc-blende structure based QWs: the circular
photogalvanic effect~\cite{PRL01}.
The crucial difference between both effects is that, while
the spin-galvanic effect may be caused by any means of spin injection,
the circular photogalvanic effect needs optical excitation with circularly
polarized radiation. Even when the spin-galvanic effect
is achieved by optical spin orientation, the microscopic mechanisms of both
effects are different.
The current flow in both the circular photogalvanic effect
and the spin-galvanic effect is driven
by an asymmetric distribution of carriers in {\boldmath$k$}-space
in systems with lifted spin degeneracy due to {\boldmath$k$}-linear terms in
the Hamiltonian. However, the spin-galvanic effect is caused by asymmetric
spin-flip scattering of spin polarized carriers and it  is
determined by the process of spin relaxation. If spin relaxation is absent,
the effect vanishes. In contrast, the circular photogalvanic effect
is  the result of
selective  photoexcitation of carriers in {\boldmath$k$}-space with circularly
polarized light due to optical selection rules.
In some optical experiments the photocurrent
may represent a sum of both effects.
For example, if we irradiate an (001)-oriented QW
by oblique incidence of circularly polarized radiation, we obtain both
selective photoexcitation of
carriers in {\boldmath$k$}-space and an in-plane component of
non-equlibrium spin polarization. Thus both effects contribute to the current
occuring in the plane of the QW.
In the  experiment  presented above we used
circularly polarized
radiation at normal incidence where the circular photogalvanic effect
is absent~\cite{PRL01} and, hence, the current is purely due to the
spin-galvanic effect.

\begin{figure}[!h]
\begin{center}
\mbox{\epsfxsize = 8cm \epsfbox{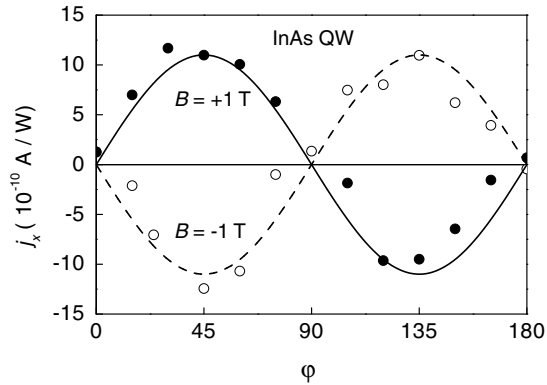}} \\
\end{center}
\caption{Magnetic field induced photocurrent  in
QWs normalized by the light power $P$ as a function of the
phase angle $\varphi$ defining the helicity for magnetic fields
of two directions.
The photocurrent  excited by normal incident radiation of
$\lambda = 148$~$\mu$m is measured
in an (001)-grown $n$-InAs QW of 15~nm width at $T$\,=4.2\,K for magnetic fields
along
$x$.      }
\label{F4}
\end{figure}

The spin-galvanic effect reported here has been obtained making use of optical
orientation of electron spins perpendicular to a QW and rotation of the spin
polarization into the QW plane by Larmor precession in an external magnetic
field. At low magnetic fields  this current phenomenologically can be also
described by a third rank tensor $\mu_{\alpha\beta\gamma}$ as
\begin{equation}
j_{\alpha} = \mu_{\alpha \beta \gamma} B_{\beta}\: i \left( {\bf E} \times
{\bf E}^* \right)_{\gamma} = \mu_{\alpha \beta \gamma} E^2 B_{\beta}
\hat{e}_{\gamma}P_{circ}\;,
\label{E1}
\end{equation}
and might also be denoted as a magnetic field induced circular photogalvanic
effect. In this equation
${\bf E}$ is the amplitude of the electric field of the
radiation, $E = |{\bf E}|$, $i \left( {\bf E} \times
{\bf E}^* \right)_{\gamma} = E^2
\hat{e}_{\gamma}P_{circ}$ and ${\bf \hat{e}}$ is a unit
vector pointing in the direction of the radiation propagation.
For $C_{2v}$ symmetry of our samples the current is described by
two independent constants and can be presented as
\begin{equation}\label{E2}
j_{x}=(\mu^\prime + \mu)E^2 B_{x} \hat{e}_{z}P_{circ} \,\,; \,\,\,
j_{y}=(\mu^\prime -\mu)E^2 B_{y} \hat{e}_{z}P_{circ} \:.
\end{equation}
The measured spin-galvanic current indeed follows the helicity of the radiation which is clearly seen in Fig.\,\ref{F4}.
%
%
\section{Summary}
In conclusion, our experimental results demonstrate that
in gyrotropic quantum wells a current occurs if electrons are
injected with an in-plane  component of  spin polarization.
Therefore the effect allows to detect spin injection into quantum wells by
measuring an electric current. Thinking on spintronic devices with
quantum wells like spin
transistors, this current must be taken into account. 
\section{\centerline{Acknowledgement}}
Financial support from the DFG, the RFFI,
the Russian Ministry of Science and the NATO
linkage  program is gratefully  acknowledged.
%

\end{document}